# Performance and Reliability Implications of Two Dimensional Shading in Monolithic Thin Film Photovoltaic Modules

Sourabh Dongaonkar[1], Chris Deline[2], and Muhammad A. Alam[1]

[1]School of Electrical and Computer Engineering, Purdue University, West Lafayette, IN

[2]National Renewable Energy Laboratory, Golden, CO

## ABSTRACT

We analyze the problem of partial shading in monolithically integrated thin film photovoltaic (TFPV) modules, and explore how the shape and size of the shadows dictate their performance and reliability. We focus on the aspects of shading problem unique to monolithic TFPV, arising from thin long rectangular series-connected cells, with partial shadows covering only a fraction of cell area. We find that due to the cell shape, the unshaded portion of a partially shaded cell experiences higher heat dissipation due to a redistribution of voltages and currents in this two dimensional geometry. We also analyze the effect of shadow size and orientation by considering several possible shading scenarios. We find that thin edge shadows can cause potentially catastrophic reverse bias damage, depending on their orientation. Finally, we show that external bypass diodes cannot protect the individual cells from shadow-induced reverse stress, but can improve the string output for larger shadows.

## 1. INTRODUCTION

As thin film photovoltaic (TFPV) technologies are moving from lab to large scale production, challenges such as process yield and long term performance reliability, which determine the levelized cost of electricity (LCOE), are becoming increasingly important [1], [2]. Therefore, the physics of solar cell degradation, and predictive modeling of long term module performance degradation is of great importance for any PV technology [3], [4]. Partial shading is one such important reliability concern for all PV modules with series connected cells. This problem was recognized early in series connected cells in crystalline PV modules used in satellites [5]. Similar problems of partial shading in terrestrial systems have also been studied actively over the years, both in terms of the module power loss [6–8], as well as reverse bias stress on shaded cells [9–11]. Consequently, different solutions such as rewiring schemes [12] and integrated bypass diodes [13] have been developed for systems using crystalline modules.

This problem of partial shading is perhaps more important for TFPV, as they are a prime candidate for building integrated and rooftop applications [14], where partial shading from nearby objects is more likely (shown schematically for a rooftop installation Figure 1(a)). The analysis of partial shading in TFPV, however, requires a tailored analysis, as their material, structure, and manufacturing differ significantly from those of the crystalline cells and modules. These differences can be summarized as follows:

- Physics of reverse breakdown in TFPV materials [15], [16] — as well as impact of moderate reverse stress [17–19] on long term performance of the modules — are very different in TFPV materials, and need to be included correctly in the analysis;

- Monolithically series-connected cells make difficult the use of integrating bypass diodes inside the modules [20], and preclude rewiring schemes available for crystalline cells;
- The long and thin rectangular shape of TFPV cells necessitates the use of two-dimensional simulations to account for the effect of shadow shape and orientation, as shown in Figure 1(b). Figure 1(b) shows the schematic of a typical TFPV module with $N_{series}$ thin rectangular series-connected cells.

In this paper, we use two-dimensional SPICE simulations (using the commercial Synopsys® HSPICE simulator) to assess the impact of different shading configurations on TFPV modules, and highlight the aspects unique to monolithic TFPV modules. The paper is arranged as follows: In Section 2, we describe the SPICE-based simulation framework, used for analyzing the shadow effects. Next, in Section 3 we explore the unique features of a shadow covering a fraction of cell area, which will be validated experimentally in Section 4. Finally, we examine the impact of various shadow sizes and orientations, and the role of bypass diodes, in Section 5.

## 2. SIMULATION FRAMEWORK

In order to mimic the actual system performance, we simulate a typical configuration of a string of TFPV modules connected to a string inverter [21], as shown in Figure 2(a). Each module is equipped with a typical power diode for external bypass [22]. The number of series connected modules is chosen (7 modules in our case) to obtain the usual string DC operating voltage of approx. 510 V [23].

Each module with dimensions $L_{module} \times W_{module}$, and $N_{series}$ series connected cells is represented using an equivalent circuit approach, as shown in Figure 2(b). We use a full two-dimensional representation for the modules by subdividing the individual cells into $N_{parallel}$ sub-cells [24], such that the sub-cell dimensions are given by $L_{sub} = L_{module}/N_{series}$ and $W_{sub} = W_{module}/N_{parallel}$. These sub-cells are connected to each other using the contact sheet resistance of the top and bottom contacts; with $R_{series} = R_{sq}L_{sub}/W_{sub}$, and $R_{parallel} = R_{sq}W_{sub}/L_{sub}$ for both top and bottom contact materials, where $R_{sq}$ is the sheet resistance of the respective contact materials. This results in 2D network-like circuit representation of the module (Figure 2(b)). Although we use single-junction a-Si:H p-i-n modules with dimensions $104 \times 120$ cm$^2$ [25] as the illustrative example, our conclusions are general and should apply to all monolithic TFPV systems including a-Si, CIGS and CdTe, once their relevant device physics is incorporated in the sub-cell equivalent circuit.

For the module dimensions considered in this work, the sub-cell area $A_{sub} = 1 cm^2$. We represent the sub-cell using the equivalent circuit of a-Si:H solar cell, including the generation enhanced recombination $J_{rec,ph}$ [26], voltage dependent collection $J_{photo}$ [27], which are functions of the total absorbed photon flux $J_{abs}$; as well as dark current comprising non-Ohmic shunt current $J_{SH}$ [28], and diode current $J_D$, as shown in Figure 2(c). This physics based equivalent circuit allows us to use a-Si:H material parameters in module simulations, ensuring that the module performance is matched closely to the datasheet [25]. In our simulations, we assume that the fully illuminated regions are operating at normal 1 Sun ($\sim 100 mW/cm^2$) intensity; while the partial shade due to nearby fully blocks the directly light and only receives diffused light, with intensity ~20% of intensity of direct sunlight, so that $J_{abs,sh} = 0.2 J_{abs0}$ [29]. The exact intensity ratio between direct sunlight and shade depends on the environmental and weather conditions of a particular location. Additional details regarding the material parameters, equivalent circuit, and module parameters are provided in Appendix A.

## 3. ASYMMETRIC STRESS IN SHADED CELLS

Because of the thin, long rectangular shape of cells in a monolithic TFPV module, the module will sometimes experience shadows that cover only a fraction of the cell area (i.e., partial shadows), as shown in Figure 1(b). In order to evaluate the impact of such partial shadows, we consider a shadow at the bottom left of the module, such that $W_{sh} = W_{module}/2$, and $L_{sh} = 5cm$. Figure 3(a) shows the color plot of the photon absorption flux ($I_{abs}$) in each sub-cell in this shading scenario. For clarity, we plot only the bottom 26 cells of this module. In this situation, the photogeneration in left half of cells 1 to 5 is reduced by 80% (i.e., $L \leq L_{sh}$ and $W \leq W_{sh}$, marked region 1), while the right half ($L < L_{sh}$ and $W > W_{sh}$, marked region 2), and all other cells from 6 to 104 ($L > L_{sh}$, marked region 3) remain fully illuminated.. We perform the 2D circuit simulation for this shading scenario, assuming that all other modules in the string are fully illuminated.

We simulate this module at the string operating point to get the voltage across each sub-cell ($V_{sub}$) in the module. The results are shown in Figure 3(b) as a 2D color plot, showing that the reduced photocurrent in cells 1 to 5 have pushed them in reverse bias. Note, however, that this reverse bias voltage is essentially uniform at $V_{sub}^{shaded} \approx -8.5V < 0$ across the entire width of the cell (regions 1 and 2), while the fully illuminated cells in region 3 continue to operate in forward bias at $V_{sub}^{illum} \approx 0.8V > 0$. The sub-cell current *direction* in the partially shaded cells (regions 1 and 2), however, remains unchanged ( therefore $I_{sub}^{shaded}, I_{sub}^{illum} < 0$), as shown in the color plot in Figure 3(c). As a consequence, the shaded cells actually *dissipate* power (in our sign convention, this means $P_{sub}^{shaded} = V_{sub}^{shaded} \times I_{sub}^{shaded} > 0$), while the fully illuminated forward-biased cells in region 3 continue to produce power so that $P_{sub}^{illum} = V_{sub}^{illum} \times I_{sub}^{illum} < 0$, see

Figure 3(d). Note that this power dissipation ($P_{sub}^{shaded} > 0$) occurs in the semiconductor junction under reverse bias as $V_{sub}^{shaded} < 0$; as opposed to fully illuminated cells in forward bias, which produce useful output power ($P_{sub}^{illum} < 0$). This must not be confused with resistive power dissipation in the contact metal/TCO layers, which is much smaller and continues regardless of the presence of a shade.

This effect of partial shading causing reverse bias in shaded cells, leading to power dissipation instead of power generation, is well known and applies to all PV technologies [30]. In thin film PV, however, the thin long shape of the individual cells results in a large asymmetry in current flow through the partially shaded cell; and dictates that the unshaded region of the partially shaded cells must compensate for the loss of photocurrent in the shaded region. This is visible in the color plot of sub-cell current $I_{sub}$ in Figure 3(c); which shows that while the current through fully illuminated cells in region 3 is essentially uniform ($I_{sub}^{illum} \approx -10mA$ per sub-cell, so that for each cell with 120 sub-cells $I_{cell}^{illum} = 120 \times I_{sub}^{illum} \approx -1.2A$), there is a stark asymmetry in current flow between region 1 and region 2 of the partially shaded cells. In these cells, the current in shaded region 1 reduces so that $I_{sub}^{shaded,1} \approx -4mA$ (because of reduced photogeneration). Current continuity in the series connected cells, however, demands that the total cell current must be same, i.e., $I_{cell}^{illum} = I_{cell}^{shaded} \approx -1.2A$. Since the shaded half is supplying less current, the current through the unshaded half must increase to compensate the loss. Therefore, as seen in Figure 3(c), the current in region 2 *increases* so that $I_{sub}^{shaded,2} \approx -16mA$. This ensures that net cell current in the shaded cells $I_{cell}^{shaded} = 60 \times I_{sub}^{shaded,1} + 60 \times I_{sub}^{shaded,2} \approx -1.2A$, satisfies current continuity constraint. The most important consequence of this redistribution in $I_{sub}^{shaded}$ is that the *unshaded region 2 dissipates more power than the*

*shaded region 1*, which means $P_{sub}^{shaded,2} \approx 4 P_{sub}^{shaded,1}$ (see Figure 3(d)), because $I_{sub}^{shaded,2} \approx 4 I_{sub}^{shaded,1}$ (recall that $V_{sub}^{shaded}$ is equal in regions 1 and 2, as shown in Figure 3(b)).

Note that the transition of current flow from asymmetric distribution in the partially shaded cells, to symmetric distribution in the fully illuminated cells cannot occur instantaneously, but must be spread over several cells. As seen in Figure 3(c), the current in cells 6 to 8, which are nearest to the shaded cells, varies slightly along the width, in order to accommodate the asymmetric current entering cell 6 from the partially shaded cell 5, while maintaining current continuity. This is made possible by a rearrangement in sub-cell voltages along the width of the fully illuminated cells nearest to the shaded regions, as dictated by the sheet resistance values of the metal and TCO contacts. This asymmetry of current flow in the fully illuminated cells, however, is reduced as we move away from the shaded regions (see Figure 3(c)). This 2D rearrangement of sub-cell current and voltage, and their relation to sheet resistance, are discussed in greater detail in Section S1 in supplementary materials.

In order to clarify the origin of this asymmetric stress in partially shaded cells, we compare the IV characteristics of fully illuminated and partially shaded cells in the series-connected configuration, as shown schematically in Figure 4 for cell no. 5 (in regions 1 and 2) and 6 (in region 3). Figure 4(a) shows the IV characteristics of the shaded and unshaded halves of cell 5, and half the current of cell 6 ($I_{cell}/2$). Note that the current in region 2 is higher than $I_{cell}/2$ in order to compensate for the lower current in region 1. Moreover, the bias point of the partially shaded cell 5 ($V_{cell}^{shaded}$) is such that the total current in half cells $I_{cell/2}^{shaded,1} + I_{cell/2}^{shaded,2} = I_{cell}$ (black dashed line). The operating point of fully illuminated cell 6 also shifts slightly so that the total current in cells 5 and 6 remains continuous (solid lines in Figure 4(b)). Thus, we can see the total current output of a partially shaded cell drops; the operating points will change so that current continuity is maintained, by pushing the shaded cell in reverse bias. Moreover, the unshaded portion (region 2) will have to supply the extra current to ensure current continuity, resulting in the excess power loss in region 2.

This observation also highlights the need to differentiate between reverse voltage stress (which is uniform for regions 1 and 2), and stress due to electrical heating due to power dissipation in the junction (which is substantially higher in region 2) in partially shaded cells in TFPV modules, as apparent from Figure 3(d). Note that this 'non-local' heating due to partial shade may occur in conventional c-Si cells as well, if only part of an individual cell area is shaded. The compact shape of c-Si cells, however, ensures that the 2D effects are much less pronounced and distribution of resistive heating due to partial shade is not readily observable in thermal images with limited resolution [31], [32]. In TFPV cells, however, this prominent distinction between shaded and unshaded portions of the partially shaded cells is a consequence of the thin long cell geometry and two-dimensional current flow in TFPV modules.

## 4. EXPERIMENTAL OBSERVATION

Although monolithic structure of TFPV modules makes it difficult to access individual cells inside and make very accurate measurements, we can use noninvasive IR thermal imaging for assessing the impact of shading and evaluating the predicted behavior [32]. This imaging technique allows us to estimate the temperature in different regions of the module, in the event of shading and for building a qualitative insight into the stress behavior. The modules used for this test are Shell Eclipse-80 CIGS modules installed in a grid-tied system at NREL's outdoor test facility.  The system is composed of two series strings of seven modules each.  As a monolithically constructed module, the Shell Eclipse-80 CIGS module has long, thin cells running the length of the module. Two separate sub-modules of 42 cells apiece are integrated into a single frame, individually protected by bypass diodes and placed in series with each other. Thermal imaging of the system was conducted with a FLIR SC640 hand-held thermal digital camera. The outdoor ambient temperature was $15 - 17°C$ and incident plane-of-array irradiance was $86 - 92 mW/cm^2$ during thermal imaging.

Figure 5(a) shows the thermal image of a series-connected module under normal operating conditions, indicating a roughly uniform temperature across the module, at around 32°C. These modules are vertically oriented, as shown by the module dimensions marked in Figure 5(a). The three regions as defined in Figure 3 are also shown. For assessing the 2D shading effect, the module was shaded by partially covering few cells with a translucent cloth (27% transmittance). The shade dimensions are marked in Figure 5(b), showing that $L_{sh} \approx 19 cm$ and $W_{sh} \approx 0.75 W_{module}$. The thermal image of the modules immediately after the removal of shading fabric is shown in Figure 5(c), showing that the temperature of unshaded region 2 increased to 39–41°C, which is noticeably higher than the shaded region 1 for the partially shaded cells, where the temperature increased to 34–36°C. On the other hand, the temperature of fully illuminated cells in region 3 remains largely unchanged around 32–34°C, as expected from the simulation shown in Fig. 3. Note that the simulation predicts (asymmetric) electrical dissipation throughout the partially shaded cell (see Figure 3(d)), while the fully illuminated cells continue to *produce* power. This is also apparent in Figure 5(c), which shows that the overall temperature in fully illuminated cells (marked 3) does not change, but power dissipation in regions 1 and 2 of the partially shaded cells causes a temperature rise in these regions. Despite the local variations in temperature across the module surface due to manufacturing non-uniformities, as well as camera angle and reflections, the general trends in temperatures in regions 1, 2 and 3 are in good qualitative agreement with the simulation results.

In order to better understand the reasons behind this temperature change, we must consider the various heat fluxes on the module surface. Figure 6(a) shows a schematic of the partially shaded module showing the insolation (marked $Q_s$) and electrical (marked $Q_e$) heat flux components in different regions. The insolation heating $Q_s$ is caused by incident radiation and is proportional to the light intensity falling on the module. Since regions 2 and 3 are illuminated, we have $Q_{s2} = Q_{s3} = Q_{s0}$, where $Q_{s0}$ is equal to the fraction of incident light lost as heat under normal operation. Typically, this arises from non-radiative recombination and resistive losses in metal/TCO layers. To a good approximation we can assume $Q_{s0} \approx 0.4 \times \phi_{inc} \approx 36 mW/cm^2$ [33], where $\phi_{inc}$ is the plane-of-array irradiance which is $\approx 90 mW/cm^2$ in our case. And, due to the 27% transmittance of the cloth in region 1 $Q_{s1} = 0.27 Q_{s0} \approx 10 mW/cm^2$. From the difference in $Q_{s1}$ and $Q_{s2}$ it might appear that the asymmetric temperature rise in Figure 5(c) may be caused in part due to the difference in insolation heating.

The total heat generated during shading, however, includes heating due to power dissipation in the shaded regions as calculated in Figure 3(d) (note that electrical power flux $Q_e = P_{sub}/A_{sub}$). Extrapolating from the simulations, we can estimate $Q_{e3} \approx -10 mW/cm^2$ for the forward biased cells in region 3 which are generating output power; and $Q_{e2} \approx 3 Q_{e1} \approx 150 mW/cm^2$ for reverse biased regions 1 and 2 dissipating power for the duration of the shade. In this scenario $Q_{s2} + Q_{e2} = 186 mW/cm^2 > Q_{s1} + Q_{e1} = 60 mW/cm^2$, which is consistent with the observations in Figure 5(c).

In order to experimentally confirm that the actual contributions from insolation heating are small, we repeat the same shading experiment with IR thermal imaging while keeping the module at open circuit. At open circuit, no current flows through the module, so there is no reverse stress or electrical asymmetry ($I_{sub}(V_{OC}) = 0$ ensures that $Q_{e1} = Q_{e2} = Q_{e3}$), and we will be able to infer the differential contribution of $Q_s$ alone Comparing the thermal image of and open circuit module before shading (Figure 6(b)), to the image taken immediately after the shade is removed (Figure 6(c)); we find that the temperature difference between regions 1 and 2 is minimal, affirming that insolation alone cannot cause significant asymmetric heating.

Thus, we can be confident that the temperature difference between regions 1 and 2 observed in Figure 5(c) is indeed due to a redistribution of current flow pattern, resulting in an asymmetric electrical heating, in qualitative agreement with the theoretical predictions of the previous section. Note that the exact heating contributions from various sources will depend on cell geometry, module configuration, contact resistances, etc.; however, key prediction of the theoretical calculations regarding asymmetric heat generation in shaded cells will remain valid.

## 5. SHADOW SIZE AND ORIENTATION

So far, we have considered a shadow of specific size and orientation, and examined its impact on the shaded cells. The shape and size of shadows, however, can change due to random events, or depending on the time of the day. Fortunately, all shadows do not result in reverse bias stress across shaded cells. In order to explore the effect of different shadow sizes and shapes, we next vary the size of a shadow by changing $L_{sh}$ and $W_{sh}$, and evaluate the corresponding reverse stress as well as output power loss, under different shading scenarios. In this section, we will be focusing on the overall cell voltage $V_{cell}$ on the partially shaded cells. Therefore, we can simplify the large 2D circuit of sub-cells into a two element circuit, in which current and voltage outputs change with varying $L_{sh}$ and $W_{sh}$, respectively. The details about creating such equivalent circuit and its validation are discussed in Appendix B.

### 5.1 Reverse Voltage Stress

We first explore the relationship of shadow size with reverse bias stress on shaded cells. The top schematic in Figure 7 identifies the important shading scenarios, which are also highlighted in the plots. Figure 7(a) shows the worst-case reverse voltage, as a function of shadow width and length. Each point on the x-y plane represents a shadow of certain width and length, with the minimum $V_{cell}$ under those conditions displayed on z-axis (also the corresponding color). In the absence of shadows, all cells operate at their maximum power point ($\sim 0.8V$), as expected (marked 1). The importance of shadow orientation can be illustrated by considering the case of thin wide shadows along the edges. Note that for asymmetric edge shadows, where a few cells are shaded fully along the width (marked 3), the shaded cells can go into severe reverse bias and breakdown ($\sim -10V$). On the other hand, a similar shadow oriented along the vertical edge is completely harmless, since all the cells are shaded equally and stay in forward bias after shading (marked 2). Most typical shadows, however, will be smaller than the panel dimensions, and would result in moderate (-2 to -5V) reverse bias stress on shaded cells (marked 4). This orientation dependence of partial shading validates the related module installation instructions from TFPV manufacturers [34].

### 5.2 Output Power Loss

We can also evaluate the impact of various shading scenarios on string DC output power loss. The plot in Figure 7(b) shows the DC output power of the string, for a variety of different shadow sizes. We can see that large shadows cause significant output power loss, but do not cause extreme reverse stress on individual cells, because the reverse bias is distributed across many shaded cells. On the other hand, the power loss is small for edge shadows, but the asymmetric edge shadow causes the worst reverse stress. This underscores the fact that loss of output power may not correlate with large reverse stress on shaded cells, as large shadows cause power loss but do not cause large stresses. This also brings into focus the role of external bypass diodes, and their ability to prevent output loss and/or reverse stress.

### 5.3 Role of Bypass Diodes

In order to gauge the role of external bypass diodes (as shown in Figure 2(a)), we perform the shading simulations for all shadow sizes, with and without external bypass. Figure 8(a) plots the difference in worst case shaded cell voltages, as a result of an external bypass diode, i.e. $\Delta V_{cell}^{bypass} = V_{cell}^{bypass} - V_{cell}^{no,bypass}$. We find that for a few shading scenarios, the reverse stress voltages are smaller in presence of external bypass. This improvement however, is small ($< 1.5V$) in the best of cases. Moreover, it shows that the external bypass diodes become active only for large shadows, when the module operating voltage becomes negative. This means that for cases like asymmetric edge shadow, which causes reverse breakdown in shaded cells (see figure 5(a)), the bypass diodes remain off. Therefore, external bypass diodes cannot provide any significant protection to individual cells from shadow stress.

External bypass diodes do, however, help in alleviating the output power loss, as seen by plotting the difference in string output power in presence of external bypass ($\Delta P_{out}^{bypass} = P_{out}^{bypass} - P_{out}^{no,bypass}$). As shown in Figure 8(b), we can get up to $200W$ extra power output from the string, when modules have external bypass diodes, but this improvement is only obtained for large area shadows. Thus, we see that external bypass diodes cannot protect individual cells from shadow-induced reverse stress and damage, but alleviate power loss in the case of large shadows on the panel.

## 6. SUMMARY AND CONCLUSIONS

In summary, we have explored the issue of partial shading in monolithic TFPV modules, using full 2D circuit simulations. We show that in order to correctly evaluate the reliability and performance implications of partial shading, the cell shape, as well as the shadow size and orientation must be taken into account. Based on our simulations we have identified some unique aspects of partial shading in these TFPV modules. These can be summarized as:

- For shadows covering only a fraction of a cell, the reverse stress voltage is distributed uniformly across the entire cell. The unshaded portion of the partially shaded cell, however, carries more current and dissipates more power, as a consequence of the 2D nature of shading and the unique TFPV cell geometry.
- The size and orientation of partial shade has a significant impact on the level of reverse stress. While thin and wide asymmetric shadows result in the worst reverse bias stress, symmetric shading causes no reverse stress at all.
- The output power loss depends on the total shadow size, and high output power loss does not imply worst-case shadow stress, and vice versa.
- External bypass diodes turn on only when the total module voltage < 0. This means that while they can improve string output power for large shadows, bypass diodes have little to no impact on the prevention of shadow-induced reverse stress.

Although the calculations here were conducted with a-Si:H p-i-n modules of certain size and configuration in mind, the conclusions are generally true for all monolithic TFPV modules, regardless of materials or processes involved. The results stem primarily from the geometry of the cell configuration and the lack of bypass diodes across individual cells.

**APPENDIX A**

In the simulations we have used a-Si:H p-i-n solar cells, and modules as our baseline. The various current components in the physical equivalent circuit of a-Si:H depend on different parameters. The voltage-dependent collection of carriers gives rise to the following voltage dependent current source [27],

$$J_{photo} = \underbrace{qGd}_{J_{abs}} \left[ \coth\left(\frac{q(V - V_{bi})}{2k_B T}\right) - \frac{2k_B T}{V - V_{bi}} \right], \tag{a1}$$

where, $G$ is effective photogeneration rate in $cm^{-3} s^{-1}$, $d$ is i-layer thickness, $V_{bi}$ is the built-in potential, $q$ is electron charge, $k_B$ is Boltzmann's constant, and $T$ is absolute temperature. The bulk recombination is also enhanced by photogeneration [26], and for the p-i-n cells is given as,

$$J_{rec,ph} \approx \underbrace{qGd}_{J_{abs}} \frac{d^2}{4(\mu\tau)_{eff}(V_{bi} - V)}, \tag{a2}$$

where $(\mu\tau)_{eff}$ is the effective mobility-lifetime product for the material. Both these generation dependent terms are a function of the absorption current $J_{abs} = qGd$, which will be modified in the partially shaded areas as $J_{abs,sh} = J_{abs0}/5$ [29].

The dark current in a-Si:H cells is a combination a diode current $J_D$, in parallel with a non-Ohmic parasitic shunt current $J_{SH}$. The diode current is given by the familiar exponential relationship, with ideality factor close to 2 for high bulk recombination, and is given as,

$$J_D \approx q\frac{n_i}{\tau_{eff}}d\frac{2k_BT/q}{V_{bi}-V}\exp\left(\frac{qV}{2k_BT}\right), \tag{a3}$$

where $\tau_{eff}$ is the effective carrier lifetime, and $n_i$ is the intrinsic carrier density. Finally, the parasitic space-charge-limited shunt current [28] has a power law voltage dependence given by,

$$J_{SH} \approx G_{SH0}V^{n_{sh}}. \tag{a4}$$

The shunt power exponent $n_{sh}$ is close to 2.5 for a-Si:H cells. For cells with a given area, all current terms will be multiplied by cell area, and the series resistance is calculated from the sheet resistance and contact cross section [24].

We simulate a typical commercial a-Si:H module with aperture $104 \times 120$ cm², and 104 cells in series [25]. The material properties in the equivalent circuit were chosen to match the module output characteristics, and are summarized in Table A1.

**APPENDIX B**

In our simulations we have assumed that all the sub-cells have identical IV characteristics, representing the average performance. Therefore, we can simplify the circuit representation of a module and reduce it to a single equivalent circuit. For a module with $N_{series}$ cells in series each with area $A_{cell}$, and module dimensions $W_{module} \times L_{module}$ (as shown in Figure 1(b)), we can use a single equivalent circuit by appropriately scaling the terminal voltage, current and series resistance; so that $V_{eq} = V_{terminal}/N_{series}$, $I_{eq} = A_{cell}J_{terminal}$, where $J_{terminal}$ is the sum of all current components in Figure 2(c), and $R_{s,eq} = R_{sq}L_{module}/W_{module}$. These scaling relations are a straightforward consequence of series (addition of voltage) and parallel (addition of current) connection of individual sub-cells.

Similar simplification is possible for a partially shaded module, which can be divided into 3 equivalent circuits connected in series and parallel, as shown in Figure B1. Based on the number of series connected cells in the region, the area of cells in the region, and the length and width of the region under consideration, we can use the same voltage, current and series resistance scaling as above (details in Figure B1). We also need to include a parallel connection between the shaded and unshaded fraction at the bottom. This needs a parallel resistance from the midpoint of the shaded region to the midpoint of unshaded region on the right, as shown in Figure B1.

Once the equivalent circuits and interconnections are set up with correct scale parameters, we can compare its output to the full 2D circuit simulation with partial shading. For the shading conditions considered in Section 2, we plot the results obtained from the full and simplified simulations in Figure B2. We find that the string level characteristics (Figure B2(a)), as well as the module level characteristics for shaded and unshaded modules (Figure B2(b)) are absolutely identical for 2D and 1D simulations. The characteristics across the shaded and unshaded halves are also the same (Figure B2(c)), as are the operating points for the two simulations. This comparison shows that the simplified equivalent circuit approach does not compromise on the accuracy of the simulations, and allows us to sweep the shadow sizes at a much smaller computational expense.

**ACKNOWLEDGMENT**

This work was supported by Semiconductor Research Corporation – Energy Research Initiative (SRC-ERI), Network for Photovoltaic Technology (NPT). The computational resources were provided by Network for Computational Nanotechnology (NCN). We would like to thank Patrick McCarthy for this help with IR imaging.

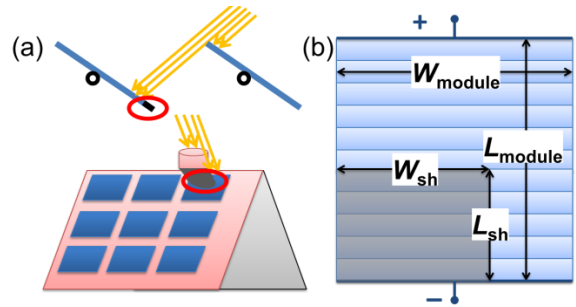

Figure 1. (a) Schematics showing possible shading scenarios due to neighboring panels or other nearby objects. (b) Schematic of a typical TFPV module, with active area dimensions, and partial shadow covering a fraction of some cells at the bottom.

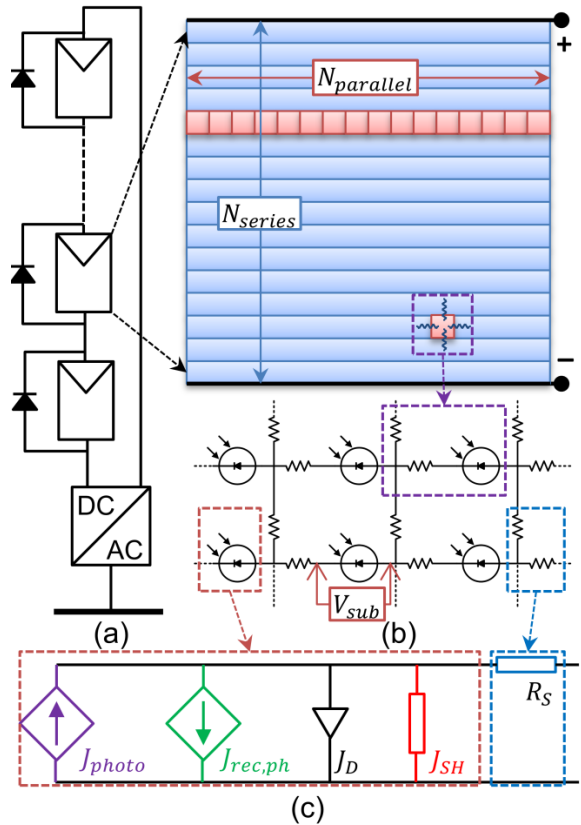

Figure 2. (a) Typical series connected string of modules (with external bypass diodes) with a string inverter. Number of modules is varied to obtain the string operating voltage of about 510V DC. (b) Each module in the string is represented as a 2D network of small sub-cells (red squares), connected through the contact sheet resistances (magenta). The sub-cell (brown box) and sheet resistance (blue box) components are marked, and the sub-cell voltage $V_{sub}$ is defined as shown. (c) In this study, we use a physics based equivalent circuit for the a-Si:H sub-cells, with correct voltage dependent photo and dark current components.

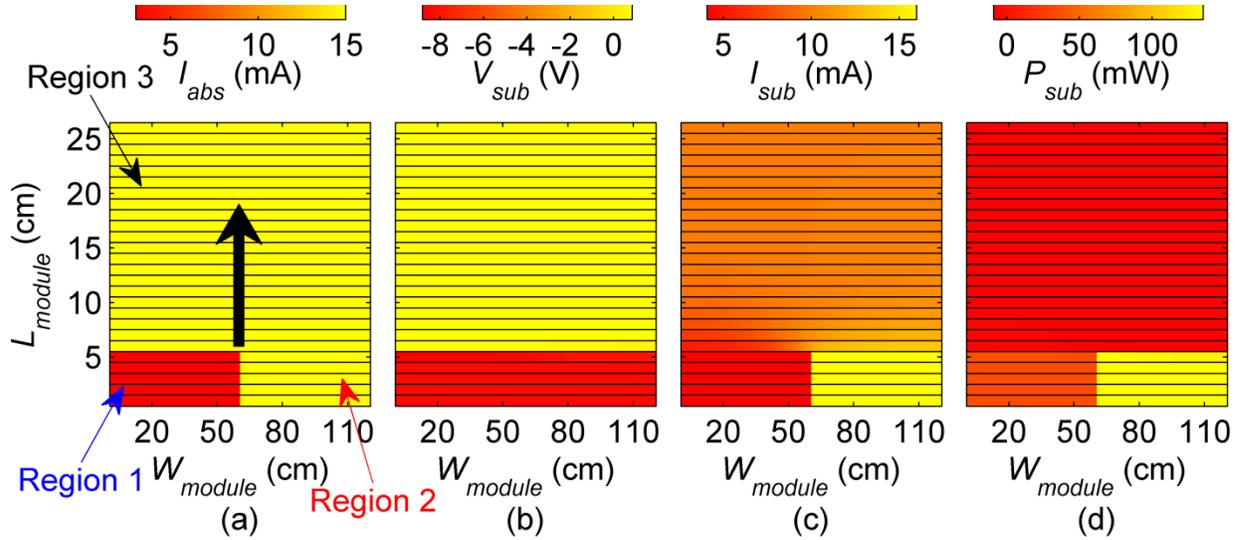

Figure 3. 2D color plots of 26 rectangular cells (y-axis scaled for clarity) in a TFPV module with a partial shadow covering left half of 5 cells at the bottom (arrow shows direction of current flow, and the color bar on the top denotes the values). (a) The photo-generated current $I_{abs}$ in sub-cells is reduced for left half of cells 1 to 5 (marked region 1), while their right half (marked region 2), and cells 6 and above (marked region 3) are fully illuminated. (b) The distribution of sub-cell voltage ($V_{sub}$) in this scenario shows that all sub-cells in the 5 partially shaded cells are reverse biased with $V_{sub}^{shaded} \approx -8.5V$ across the cell width (regions 1 and 2), but the fully illuminated cells continue to operate in forward bias so that $V_{sub}^{illum} \approx 0.8V$. (c) The sub-cells current $I_{sub}$ shows asymmetric behavior, and $I_{sub}$ in region 2 increases to $\sim -16mA$, to compensate for the low $I_{sub} \approx -4mA$ in region 1, to ensure current continuity with fully illuminated cell where $I_{sub}^{illum} \approx -10mA$. (d) As a consequence of the voltage and current redistribution in cells 1 to 5, the sub-cell power dissipation $P_{sub}$ in region 2 is four times that of region 1 (positive values in this sign convention), while the fully illuminated cells 6 and above continue to produce power (negative values in this sign convention).

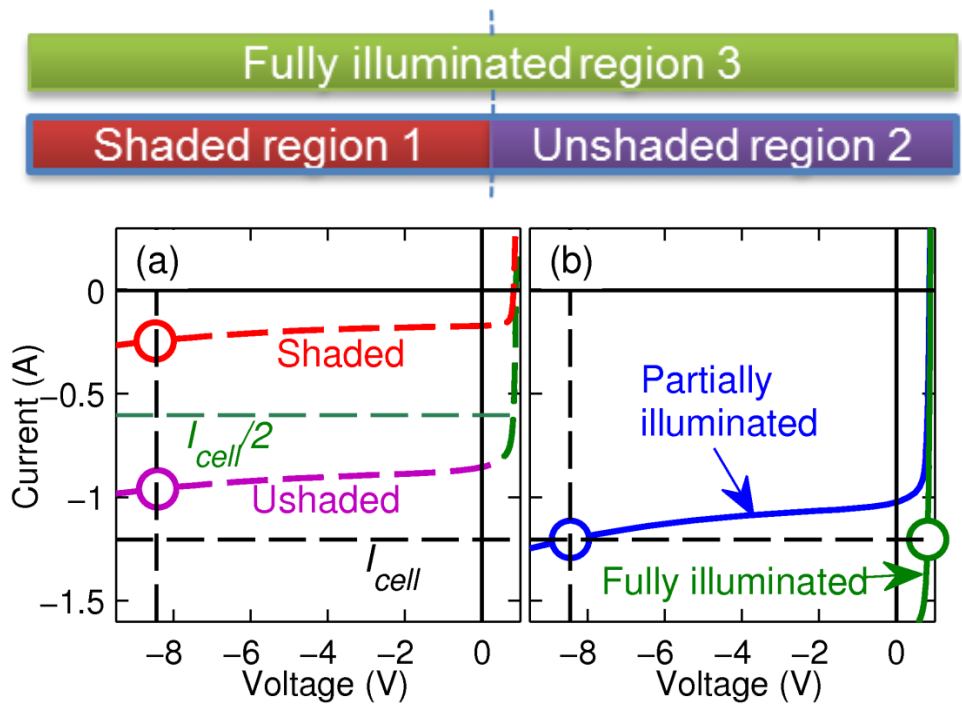

Figure 4. A schematic showing a pair of cells at the edge of the shaded and unshaded regions in Fig. 3 namely, fully illuminated cell 6 in region 3 (green) in series with the partially shaded cell 5 (blue), with a shaded region 1 (red) and unshaded region 2 (magenta). (a) The IV curves of half cells, including the shaded (red), unshaded (magenta), and fully illuminated (green) halves, show that as the current in region 1 drops the current in region 2 must increase beyond even $I_{cell/2}$ to maintain current continuity. (b) The full IV curves of the cells (solid lines) showing that the total current through partially shaded cell (blue) is same as the full illuminated cell (green), and their respective operating voltages determined by the amount of photocurrent reduction in the shaded cell.

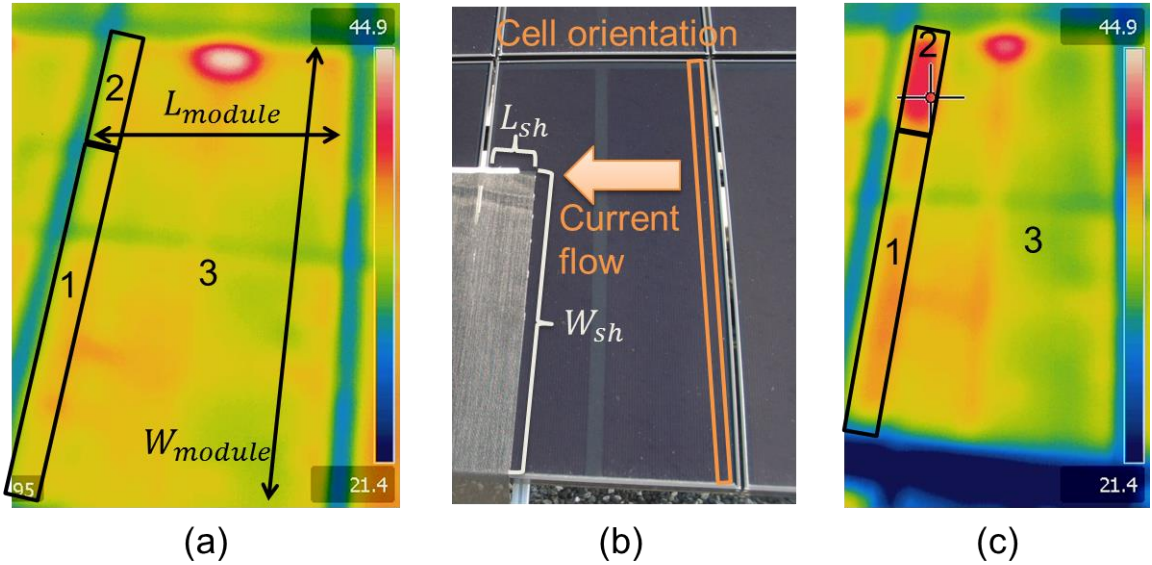

Figure 5. (a) IR thermal image of fully illuminated series connected modules under normal operating conditions showing roughly uniform operating temperature of 32–34°C (color bar on the right indicates temperature in °C). The hot spot at the top appears in all modules, due to placement of the connectors in that region causes current crowding and local heating. The module dimensions on this vertically oriented modules, and positions of regions 1, 2, 3 as defined in Figure 3 are also shown. (b) An image of the same module under partial shade using a translucent shading cloth with 27% transmittance. The shade dimensions, direction of current flow and cell orientation are also marked. (c) The IR thermal image immediately after removal of the shade showing a slight temperature rise to 34–36 °C shaded region 1, but the temperature in unshaded region 2 is noticeably higher at 39-41 °C, and the temperature of fully illuminated region 3 is largely unchanged, as anticipated by theory.

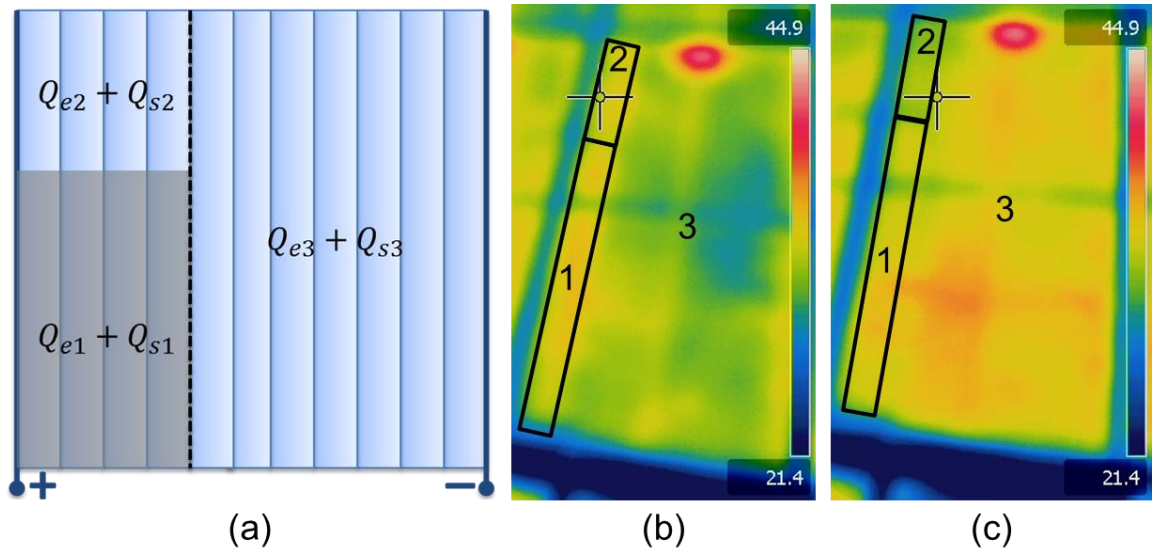

Figure 6. (a) Schematic of partially shaded module showing the electrical ($Q_e$) and insolation heat ($Q_s$) flux components in the three regions of interest. (b) IR thermal image of module in steady state at $V_{OC}$ showing roughly uniform temperature across the module surface, with relatively small extrinsic variation. This is then subjected to the identical shading conditions as in Figure 5. (c) IR thermal image immediately after the partial shade is removed does not show any significant difference in temperatures between regions 1, 2, and 3; demonstrating that the differential heading due to insolation is essentially negligible.

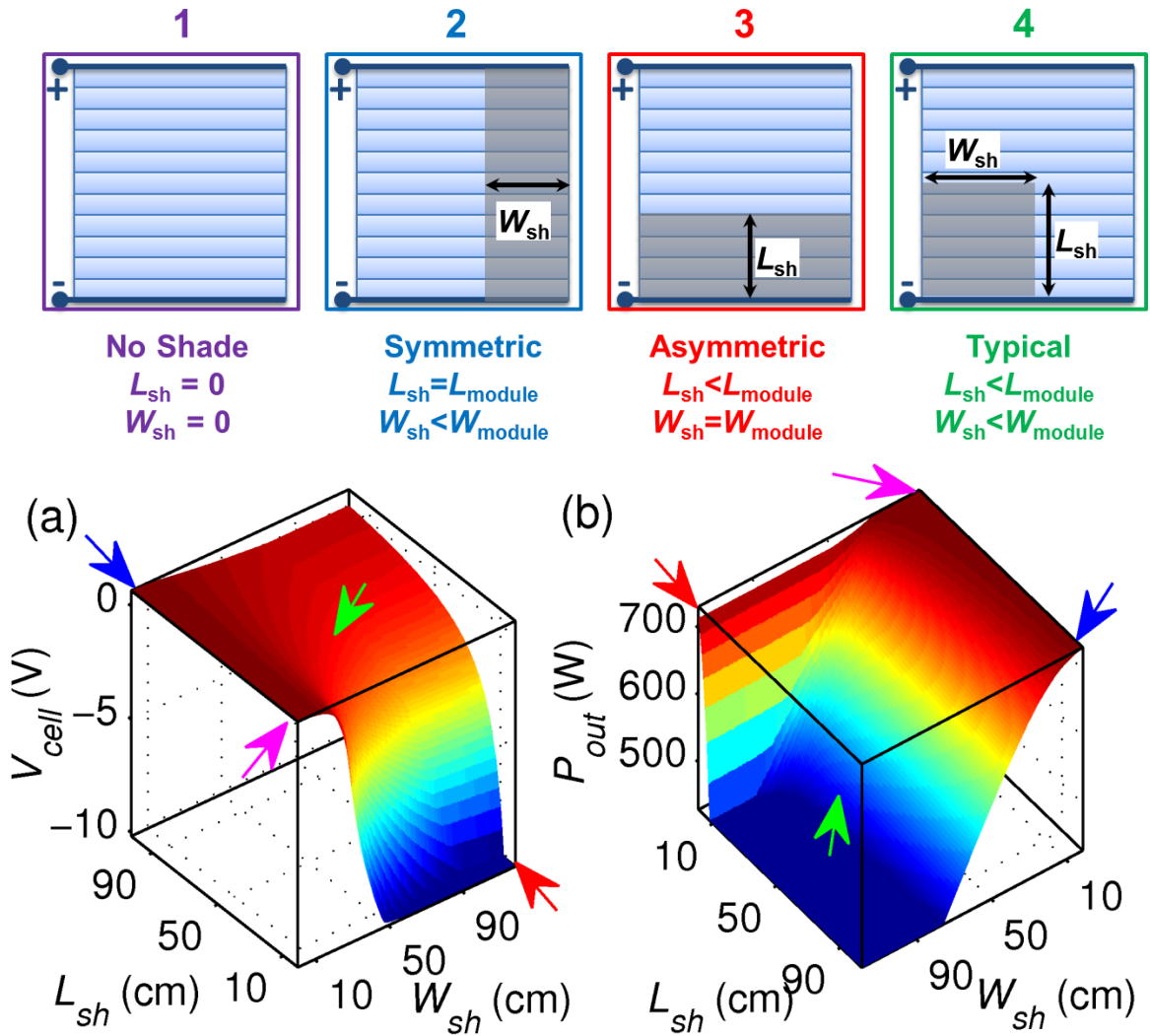

Figure 7. Schematics at the top highlight the important shading scenarios, with consequences shown in plots below. (a) Surface plot showing the worst case reverse bias (z-axis) across a cell in the shaded region for shadows of different length (y-axis) and width (x-axis). Note that unlike the symmetric edge shading (marked 2), the asymmetric edge shadow (marked 3) causes reverse breakdown of shaded cells, but typical shadow stresses (marked 4) are fairly moderate. (b) String output power (z-axis) as a function of varying shadow sizes, shows that small edge shadows (2 and 3) do not cause significant drop in output power, which only happens for large shadow sizes.

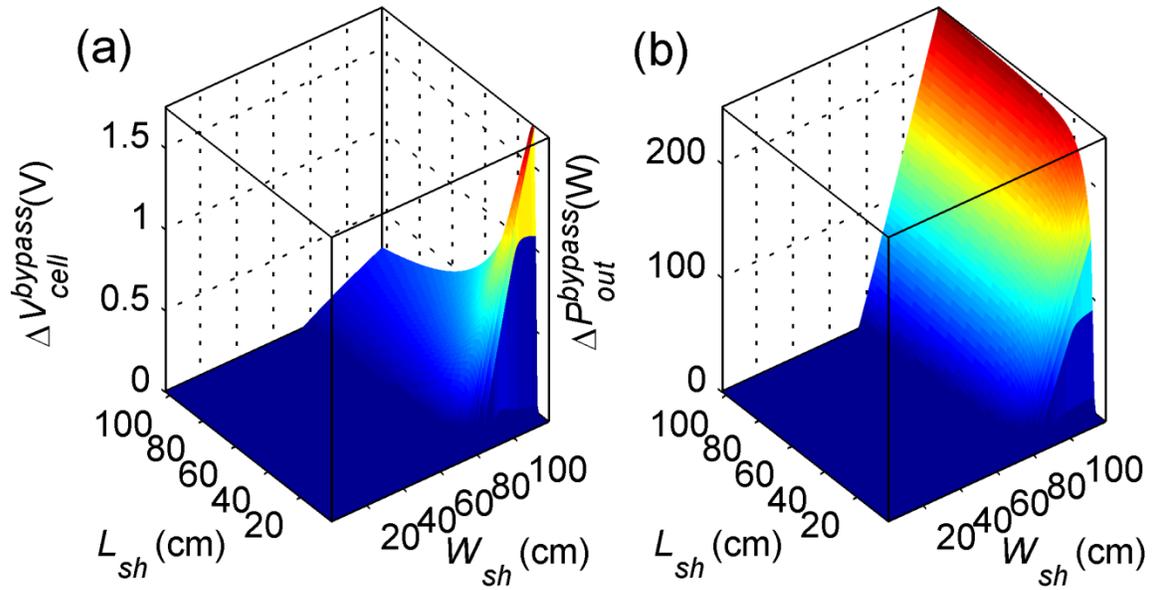

Figure 8. Role of external bypass diodes is explored by comparing the cases with or without bypass diodes. (a) $\Delta V_{cell}^{bypass}$ is the difference between worst case reverse stresses (z-axis) in presence and absence of external bypass is plotted for different shadow sizes, showing that external bypass only marginally (by about 1V) reduces the shadow induced stress, typically for larger shadows. (b) The difference in power output $\Delta P_{out}^{bypass}$, due to external diodes (z-axis) shows that while there is no impact for smaller shadows, for large shadows external bypass diodes can improve string DC output by about $200W$.

| Module parameters | |
| --- | --- |
| $L_{module} \times W_{module}$ | 104 cm × 120 cm |
| $N_{series}, N_{parallel}$ | 104, 120 |
| Material parameters | |
| $J_{abs0}$ | $15 \: mA/cm^2$ |
| $V_{bi}$ | $1.1 \: eV$ |
| $(\mu\tau)_{eff}$ | $10^{-8} \: cm^2/V$ |
| $G_{SH0}, n_{sh}$ | $5 \: \mu A/cm^2, 2.5$ |
| $R_{sq}^{TCO}, R_{sq}^{Metal}$ | $10 \Omega/sq, 0.1 \Omega/sq$ |
| $V_{BD}$ | $-10V$ |

Table A1. Module dimensions and material parameters used in these simulations to accurately represent the cell and module level characteristics of a-Si:H p-i-n solar cells.

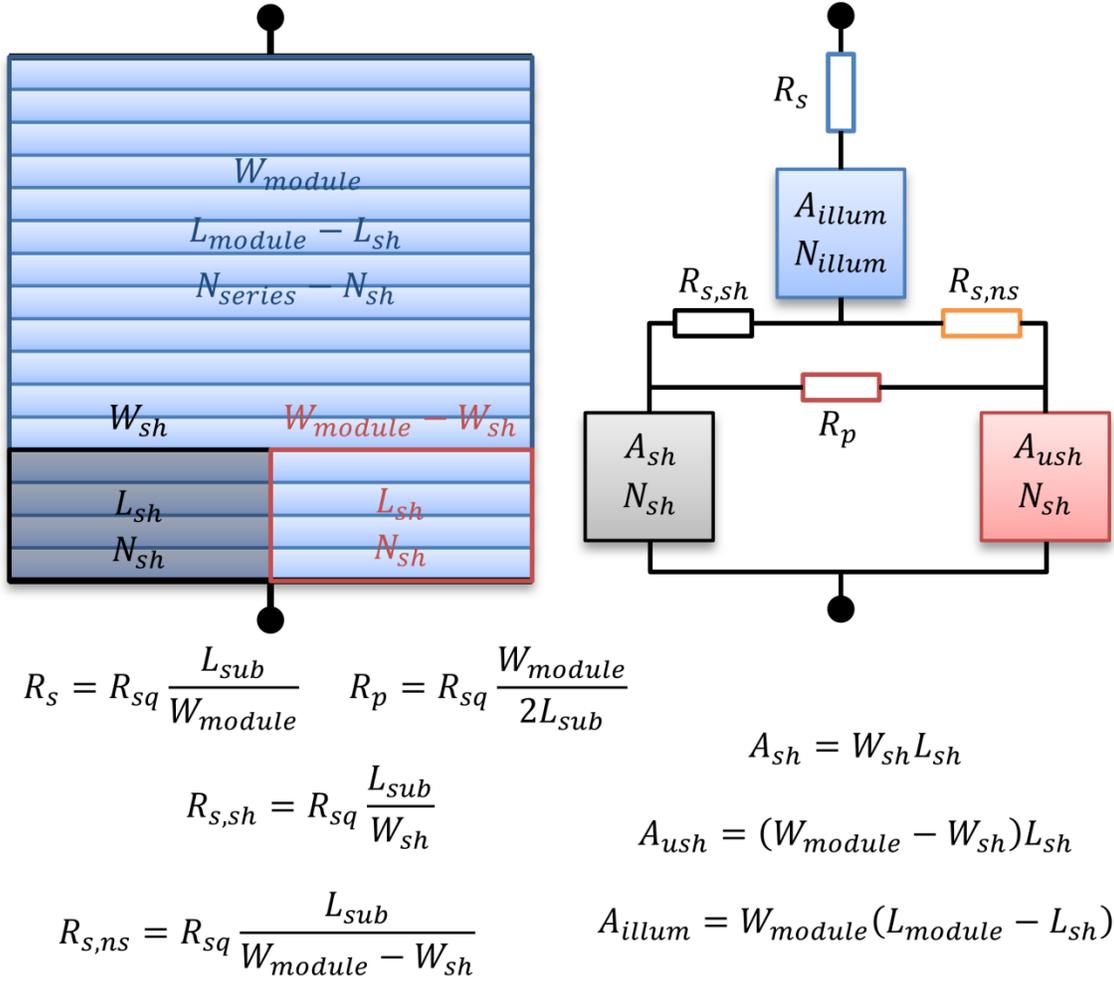

Figure B1. Schematic showing a partially shaded module with module and shadow dimensions, along with the simplified 3-element equivalent circuit with blocks for the shaded and unshaded fractions of the partially shaded cells, and fully illuminated cells of the module. The current and voltage scaling parameters, as well as equivalent resistance values are shown.

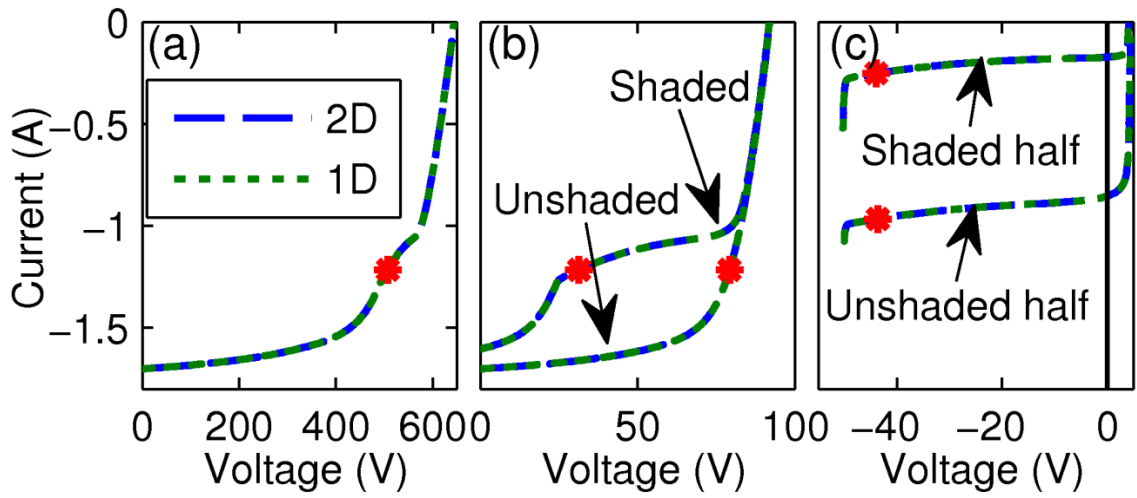

Figure B2. IV characteristics comparing the full 2D simulations (blue dashes) with the simplified 1D case (green dotted) for the partial shading scenario considered in Section 2. The 1D and 2D cases are virtually indistinguishable, showing that the (a) string output, (b) module output for shaded and unshaded modules, as well as (c) characteristics of the shaded and unshaded halves of the partially shaded cells, are identical.